\documentclass[twocolumn,letterpaper,superscriptaddress,floatfix]{revtex4-1}
\usepackage[space]{grffile}
\usepackage{mathrsfs}
\usepackage{bm,graphicx,graphics,amsmath,amssymb,bm,epsfig,color}
\usepackage{euscript,multirow,tabularx}
\usepackage{longtable}
\usepackage{pifont}
\usepackage{xcolor}
\usepackage{float}
\usepackage{changes}
\usepackage{ulem}
\usepackage{lineno}
\graphicspath{{./Figs/final/}}
\usepackage{lipsum}
\begin{document}

\title{Optimizing the magnon-phonon cooperativity in planar geometries}

\author{K. An}
\affiliation{Quantum Technology Institute, Korea Research Institute of Standards and Science, Daejeon 34113, Republic of Korea}

\author{C. Kim}
\affiliation{Quantum Technology Institute, Korea Research Institute of Standards and Science, Daejeon 34113, Republic of Korea}

\author{K.-W. Moon}
\affiliation{Quantum Technology Institute, Korea Research Institute of Standards and Science, Daejeon 34113, Republic of Korea}

\author{R. Kohno}
\affiliation{Université Grenoble Alpes, CEA, CNRS, Grenoble INP, Spintec, 38054 Grenoble, France}

\author{G. Olivetti}
\affiliation{Université Grenoble Alpes, CEA, CNRS, Grenoble INP, Spintec, 38054 Grenoble, France}

\author{G. de Loubens} 
\affiliation{SPEC, CEA-Saclay, CNRS, Université Paris-Saclay, 91191 Gif-sur-Yvette, France}
  
\author{N. Vukadinovic}
\affiliation{Dassault Aviation, 92552 Saint-Cloud, France}

\author{J. Ben Youssef} 
\affiliation{LabSTICC, CNRS, Universit\'e de Bretagne Occidentale, 29238 Brest, France}

\author{C. Hwang}
\email[Corresponding author: ]{cyhwang@kriss.re.kr}
\affiliation{Quantum Technology Institute, Korea Research Institute of Standards and Science, Daejeon 34113, Republic of Korea}

\author{O. Klein}
\email[Corresponding author: ]{oklien@cea.fr}
\affiliation{Université Grenoble Alpes, CEA, CNRS, Grenoble INP, Spintec, 38054 Grenoble, France}

\date{\today}

\begin{abstract}

Optimizing the cooperativity between two distinct particles is an important feature of quantum information processing. Of particular interest is the coupling between spin and phonon, which allows for integrated long range communication between gates operating at GHz frequency. Using local light scattering, we show that, in magnetic planar geometries, this attribute can be tuned by adjusting the orientation and strength of an external magnetic field. The coupling strength is enhanced by about a factor of 2 for the out-of-plane magnetized geometry where the Kittel mode is coupled to circularly polarized phonons, compared to the in-plane one where it couples to linearly polarized phonons. We also show that the overlap between magnon and phonon is maximized by matching the Kittel frequency with an acoustic resonance that satisfies the half-wave plate condition across the magnetic film thickness. Taking the frequency dependence of the damping into account, a maximum cooperativity of about 6 is reached in garnets for the normal configuration near 5.5 GHz.
\end{abstract}

\maketitle

Quantum information processing relies on the coherent interconversion process between distinct particles. Among the candidates, the coupling between magnons and phonons offers a certain advantage in terms of tunability and long-range propagation distance, which make them useful for building an efficient quantum transducer \cite{lachance2019hybrid}. Magnetically driven circularly polarized phonons can also carry angular momentum \cite{zhang2014angular,garanin2015angular} and transfer it over a characteristic distance much longer than that of magnons without requiring a magnetic material \cite{an2020coherent,schlitz2022magnetization,ruckriegel2020long}. This new direction in spintronics has created a surge of interest in using phonon angular momentum to enable long-distance spin transport \cite{brataas2020spin,li2021advances}. Furthermore, phonons are tunable via the geometry and size, raising the hope of optimizing the conversion process \cite{mondal2018hybrid,berk2019strongly,godejohann2020magnon}.

The mutual coupling can be interpreted as an interaction between lattice displacement and magnetic orientation via magnetostriction \cite{kittel1958interaction,spencer1958magnetoacoustic,schlomann1960generation,matthews1962acoustic,lee1955magnetostriction} or microscopically in terms of magnon-phonon interaction \cite{guerreiro2015magnon,holanda2018detecting,rezende2021theory}. The coupling results in the formation of \textit{magnon-polaron} hybrid states \cite{shen2015laser} that are split by the coupling strength, $\Omega$. Experimentally strong coupling has been demonstrated in various systems such as magnetic insulators \cite{zhang2016cavity,khivintsev2018magnetoelastic}, metals \cite{zhao2020phonon}, semiconductors \cite{kuszewski2018optical}, and nano-fabricated ferromagnets \cite{berk2019strongly,godejohann2020magnon}.

\begin{figure}[b]
\centering
\includegraphics[width=0.48\textwidth]{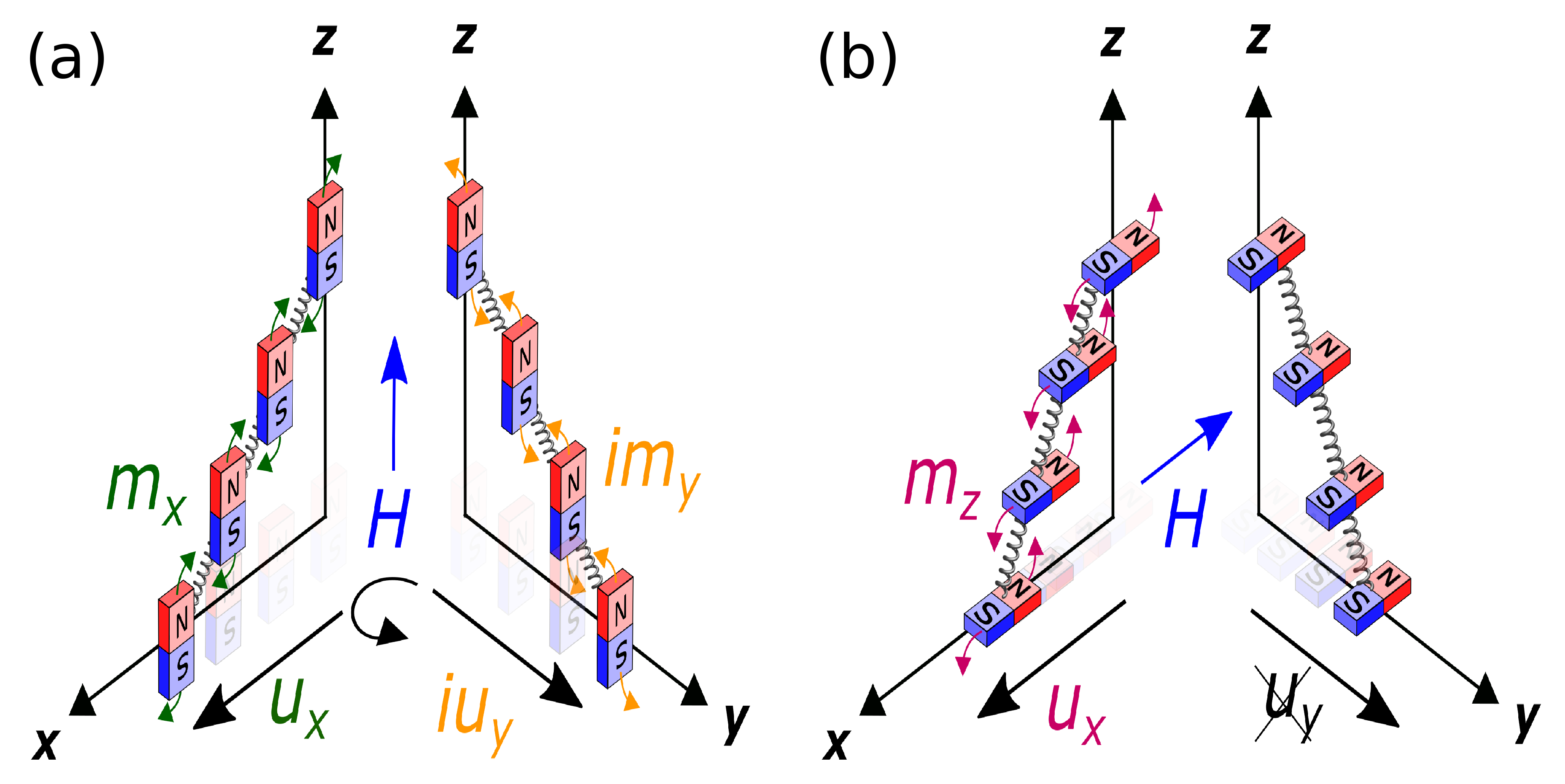}
\caption{(Color online) Schematic illustration of the coupled dynamics between spins and lattice in a thin magnetic film magnetized either along (a) the out-of-plane or (b) in-plane directions. The magnetoelastic coupling, $\Omega$, is proportional to the dot product between the oscillating components of the magnetization, $m_i$, and the vertical gradient of lattice displacement, $\partial_z u_i$. In (a), both $m_x$ and its 90 degree rotated component $im_y$ couple to $\partial_z u_x$ and $i\partial_z u_y$. Hence the Kittel mode couples to circularly polarized phonons in the $xy$-plane and both transverse components contribute to $\Omega$. In (b), $\partial_z u_y$ does not couple to the dynamic magnetization and only $m_z$ couples to $\partial_z u_x$, leading to the excitation of linearly polarized phonons. The lack of coupling with $u_y$ leads to a reduction in $\Omega$.}
\label{FIG_illlus}
\end{figure}

To gain control over the interconversion process, one needs to develop a way to tune the cooperativity $\mathscr{C}=\Omega^2/(2\eta_m \eta_a)$ \cite{turchette1998squeezed,tuchman2006normal,kuhn2010cavity,reiserer2015cavity,al2018cooperativity,thomas2022efficient}, a figure of merit which measures the number of oscillations that occurs between the two waveforms before decoherence starts to kick in. Here $\eta_m$ and $\eta_a$ are the relaxation rates of magnons and phonons, respectively. The relaxation rates are related to the material quality issue and are challenging to control, but tuning the coupling strength can provide a more efficient way to control $\mathscr{C}$ due to its $\Omega^2$ dependence. The manipulation of coupling strength between magnons and microwave photons has been achieved by changing the position of magnetic system \cite{harder2018level,xu2019cavity,ihn2020coherent}. But this scheme is hard to apply for the magnon-phonon coupled system because two bodies are inseparable.

For magnon-phonon coupled systems, the coupling stength can be tuned by adjusting the magnetoelastic energy \cite{morgenthaler1963longitudinal,comstock1963generation}. For the normally magnetized case, the strains from both transverse lattice vibrations, $u_x$ and $u_y$, give rise to the change in magnetization alignment (Fig.~\ref{FIG_illlus}(a)). This contributes to the magnetoelastic energy  $E_{\rm me}\propto M_z (m_{x}\partial_z u_{x} + m_{y}\partial_z u_{y}$), where $M_{i}$, $m_{i}$, and $u_{i}$ represent the static, dynamical parts of the magnetization, and the lattice vibration along the $i$ direction, respectively. $z$ is the direction normal to the film surface. An analytic form of the coupling strength for this configuration is presented in Appendix A. For the in-plane magnetized case, however, only the lattice displacement along the direction of magnetic field ($u_x$ in Fig.~\ref{FIG_illlus}(b)) couples to the magnetoelastic energy, leading to $E_{\rm me}\propto M_{x} m_z\partial_z u_{x}$. This difference in $E_{\rm me}$ results in about 1.7 times weaker coupling strength for the in-plane magnetized case at $\omega=2\pi \times 6.4~\rm GHz$ (see Appendix B). Other than these two field directions, an analytic expression for the coupling strength has a complex form and a full angular dependence of coupling strength is presented in Appendix C.

The increased coupling strength for circularly polarized phonons (as in the case of out-of-plane field) is due to the simultaneous contribution of both orthogonal lattice vibrations, leading to a larger magneto-elastic energy. For the linearly polarized phonons (as in the case of in-plane magnetized case), only the component of the lattice vibration parallel to the field direction couples to the magnetization dynamics, hence leading to less effective coupling. This discrepancy accounts for an approximate two-fold enhancement in the coupling strength of circularly polarized phonons. However this enhancement factor may differ depending on the nature of phonon modes and their propagation directions relative to crystallographic axes. Furthermore, the coupling strength in thin magnetic films also depends on the overlap between the magnon and phonon  wave functions \cite{litvinenko2021tunable,schlitz2022magnetization}. This can be tuned by changing the phonon wavelength over the thin magnetic layer. For the uniform Kittel mode, the maximum coupling is achieved when the phonon half wavelength is equal to the thickness of the magnetic layer. On the other hand, the maximum coupling between spatially nonuniform magnon modes and phonons satisfies a different condition \cite{schlitz2022magnetization}. 

In our experiments, the magnon phonon coupling is detected optically by micro-Brillouin light scattering (BLS), which has the advantage of sensing magnons or phonons \textit{directly} over the laser spot size of a few microns. The local detection also reduces the contributions from spatial inhomogeneities, minimizing spectral broadening, which allows sensitive detection of coupling between quasiparticles \cite{bozhko2017bottleneck,holanda2018detecting,frey2021double}. Typical BLS experiments, however, have a limited spectral resolution on the order of ten MHz \cite{demokritov2007micro,sebastian2015micro}, which is not enough to resolve $\Omega$ of a few MHz range.

Here we show that the microwave excited BLS can overcome this limit and sense magnon-polarons with enhanced contrast at the avoided crossing. This improvement comes from reducing the adverse effects of spatial inhomogeneities. We perform experiments with both in-plane and out-of-plane magnetic field configurations. Stronger magnon-phonon coupling is observed upon applying the magnetic field to the out-of-plane direction, consistent with the illustration in Fig.~\ref{FIG_illlus}. We show that the coupling strength is further tunable via changing the phonon wavelength.


\begin{figure}
\centering
\includegraphics[width=0.48\textwidth]{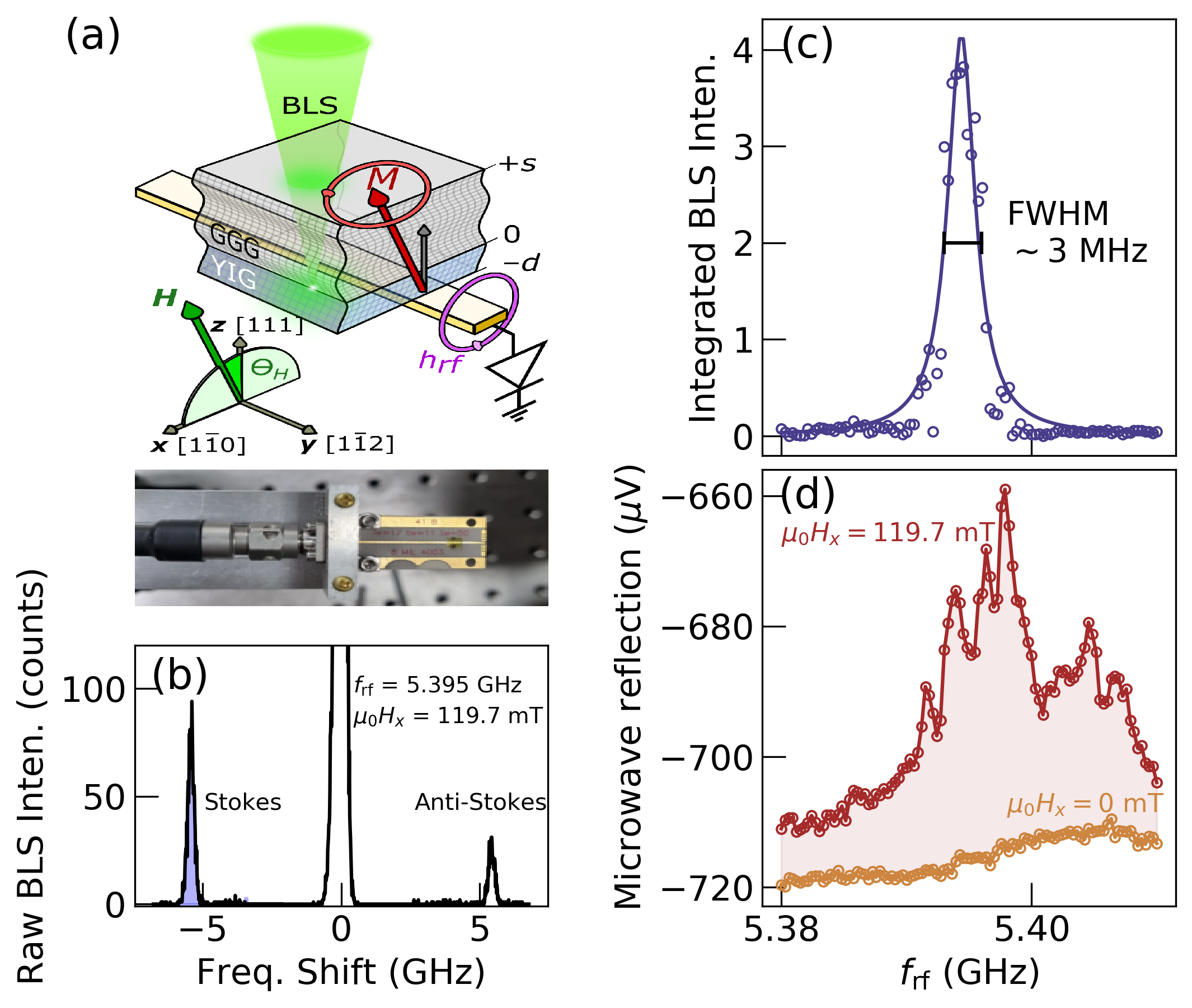}
\caption{(Color online) (a) Schematic of the experimental setup measuring simultaneously the microwave and the BLS spectra. The top panel illustrates the laser light, focused at the diffraction limit on the backside YIG layer, which is inductively coupled to a microwave antenna producing an oscillating $h_{\rm rf}$. The bottom panel shows a picture of a YIG sample placed on the antenna.  (b) Raw BLS spectrum as a function of frequency shift at a microwave power of $-15$~dBm. The colored region represents the spectral area of Stokes peak used to obtain the integrated intensity. (c) Locally integrated BLS intensity as a function of excitation frequency $f_{\rm rf}$ with in-plane magnetic field. The solid line is a Lorentzian function fit. (d) Absorption spectra simultaneously detected by the global inductive coupling to the microwave antenna. The brown and orange colors represent the spectra taken with the in-plane magnetic fields of 119.7 mT and 0 mT, respectively. The shaded area emphasizes the difference, corresponding to the absorbed microwave power by the sample. The comparison between (c) and (d) illustrates the benefit of local measurement to eliminate inhomogeneous broadening. More detailed comparison between the microwave absorption and BLS spectra as a function of magnetic field and frequency is presented in Appendix D.}
\label{FIG_setup}
\end{figure}

\begin{figure*}[t]
\centering
\includegraphics[width=1\textwidth]{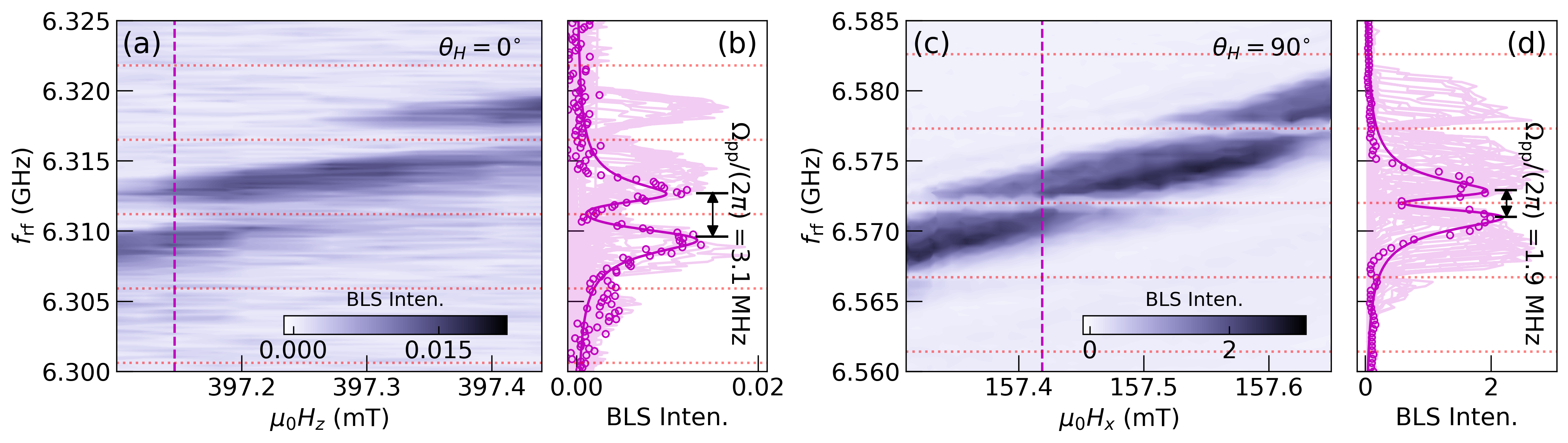}
\caption{(Color online) Comparison of the gap at the magnon-phonon avoided crossing for the (a) out-of-plane and (c) in-plane magnetic field configurations. Measurements are performed on identical frequency span. The gap $\Omega_{\rm pp}$ is clearly larger in the normal configuration ($\theta_H = 0^\circ$). The microwave power levels are $-6$ dBm and $-15$ dBm for (a) and (c), respectively \cite{footnote_fig3}. The color scales are normalized to their maximum intensities. The red dotted horizontal lines represent the phonon frequencies. (b,d) are linecuts taken from the purple dashed lines in (a,c). Solid lines represent the fitting lines using the two coupled oscillator model. The shadowed areas in (b,d) show the frequency projection of the each density plot.}
\label{FIG_compare}
\end{figure*}

We use a $d=180~\rm nm$ thick yttrium iron garnet (YIG) grown on a $s=330~\mu$m thick gadolinium gallium garnet (GGG) by the liquid phase epitaxy method, followed by removing one side of the originally double-sided YIG via ion-beam etching. The lateral size of 2 mm $\times$ 2 mm sample is placed on top of a $430~\mu$m wide strip antenna (see Fig.~\ref{FIG_setup}(a)). An amplitude-modulated microwave at a frequency of 3177 Hz is applied to the sample through a circulator to measure the microwave absorption spectrum. The reflected power is sent to a diode detector and is recorded using a lock-in amplifier. Simultaneously a 37 mW green laser is focused to a beam diameter of 4~$\mu$m at the bottom of the sample, where the YIG layer is placed, via a 20$\times$ objective lens. The beam position is fixed at about 50 $\mu$m away from the edge of antenna. The laser has a penetration depth of 6 $\mu$m in YIG \cite{scott1974absorption}, and it is transparent to GGG. As a result, it can penetrate the entire sample without any significant absorption. The $z$ component of the antenna field, $h_{\rm rf}$, is at its maximum near the edge of the antenna, where it exerts the maximum torque on magnetization with the in-plane bias field $H_x$ \cite{an2014control}. A cross-polarized back-scattered light is sent to a tandem Fabry-Perot interferometer, where the frequency shift of scattered light is analyzed. The external magnetic field can be applied along the in-plane or out-of-plane directions. A pair of secondary Helmholtz coils is used to control the magnetic field on the order of tens of $\mu$T.

\begin{figure}[ht!]
\centering
\includegraphics[width=0.48\textwidth]{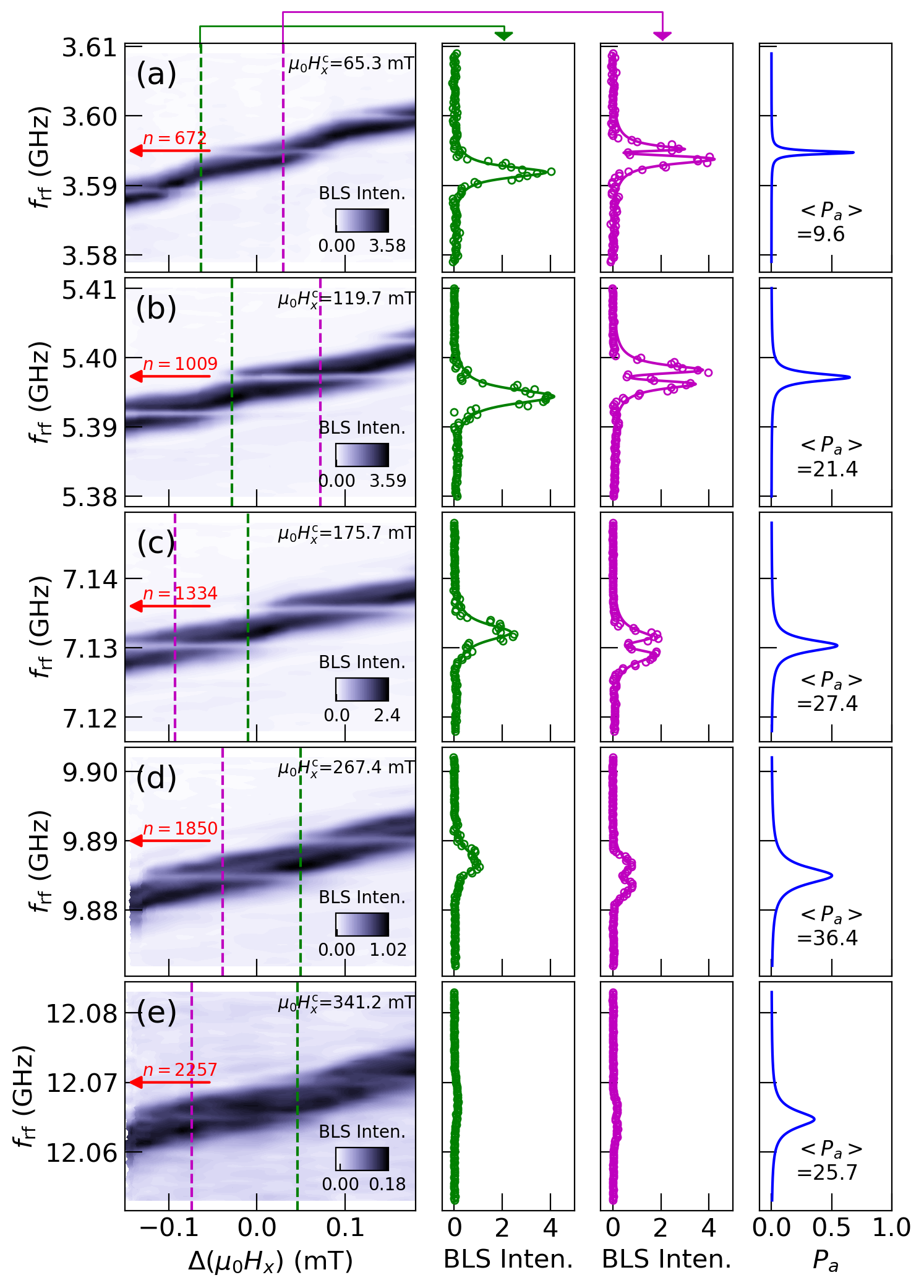}
\caption{(Color online) Magnon-phonon coupling is examined for five different amplitudes of the in-plane center field, $H_x^{\rm c}$. (a-e) Density plot of integrated BLS intensity as a function of excitation frequency and magnetic field. Color scales are normalized to the maximum intensity in each panel. Green and purple dashed lines are drawn to indicate the cases without and with magnon-phonon coupling, respectively. $\Delta (\mu_0 H_x)$ denotes the deviation from the center field $H_x^{\rm c}$. The second (third) column shows the linecuts along the green (purple) lines. Solid lines are the fits obtained from the two oscillator model. In the fourth column, we show the calculated acoustic energy dissipation. $\big< P_a \big>$ represents an integrated value over the whole spectral range, which is proportional to the coupling strength. All measurements were conducted at $-15$ dBm.}
\label{FIG_freqsweep}
\end{figure}

\begin{figure}[ht!]
\centering
\includegraphics[width=0.48\textwidth]{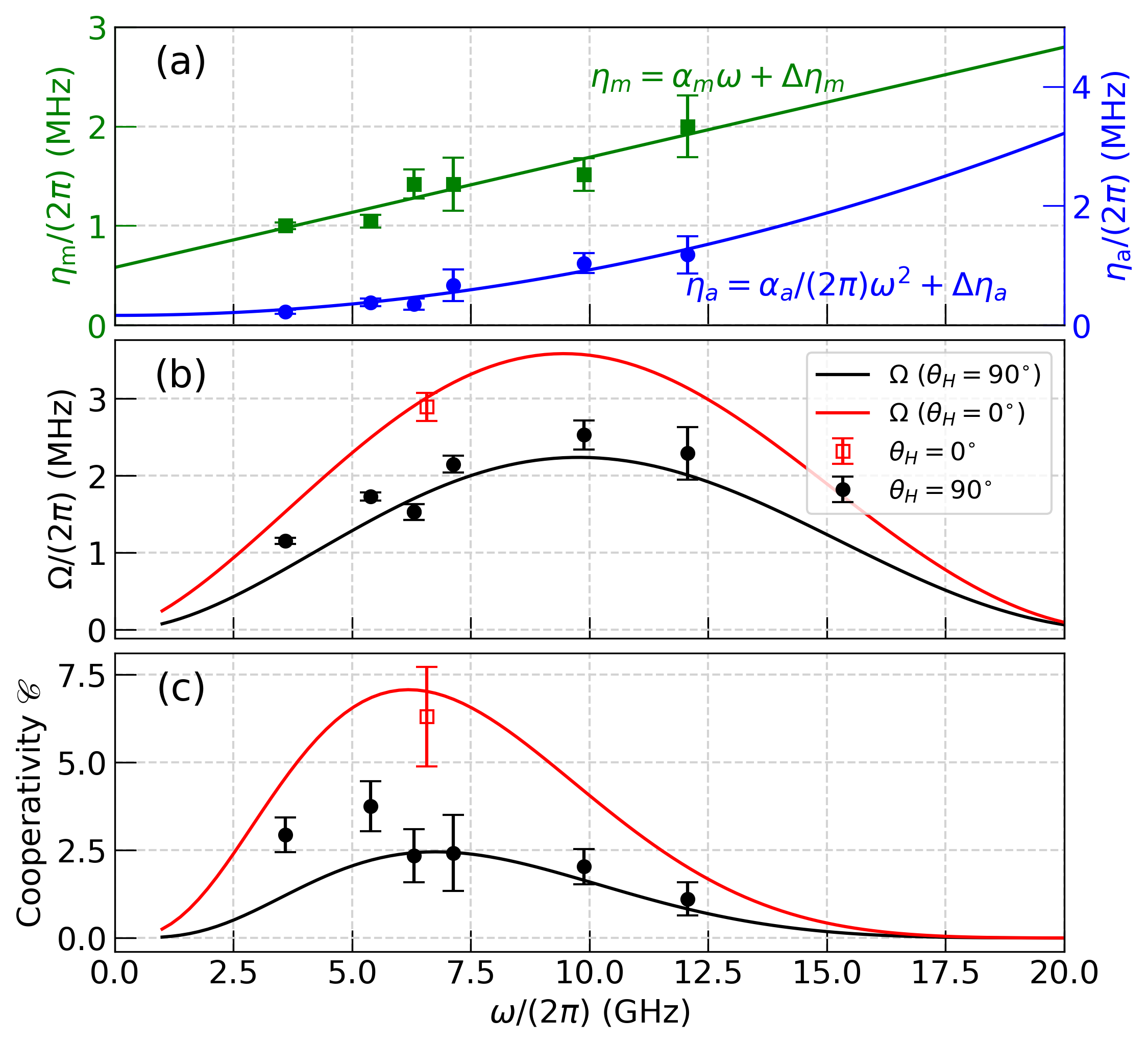}
\caption{(Color online) (a) Frequency dependence of the magnetic (green) and acoustic (blue) damping parameters. Solid lines are fit by phenomenological models. (b) Corresponding frequency variation of the magnon-phonon coupling strength evaluated from the gap at the avoided crossing. Two solid lines represent the expected behavior based on effective estimate of the material parameters (see main text). (c) Corresponding variation of cooperativity.}
\label{FIG_extract}
\end{figure}

The raw BLS spectrum under microwave excitation is shown in Fig.~\ref{FIG_setup}(b). Thermally excited magnon signals in our thin film YIG lie under the noise level. The microwave-assisted magnon signals are more substantial and typically show about 100 counts/s at the peak with an excitation power of $-15$ dBm. The two peak positions labeled by Stokes and Anti-Stokes peaks in the raw spectrum are identical to the excitation microwave frequency ($\pm f_{\rm rf}$). Here, the magnon linewidth is determined by the instrumental limit, which is about 300 MHz close to the linewidth of the peak located at zero frequency shift. The Stokes peak area is then integrated, and the integrated intensity is plotted as a function of $f_{\rm rf}$ in Fig.~\ref{FIG_setup}(c), where a lorentzian function fits the data. Here the full linewidth is about 3 MHz, comparable to the previously reported values \cite{lecraw1958ferromagnetic,beaulieu2018temperature}. The wavevector $k$ of magnons dominantly contributing to the spectrum is $k=0$ as the antenna coupling is less efficient for $k$ values larger than $\pi/w$, where $w$ is the antenna width \cite{kalinikos1980excitation}. In our backscattering geometry, the bulk standing wave phonons are not detected due to wavevector mismatch \footnote{The bulk phonon wave vector that satisfies the momentum conservation for our backscattering geometry is given by $k_s-k_i = 4\pi n/\lambda$, where $k_s$ and $k_i$ are the scattered and incident light wave vectors. The refractive index $n$ is about 2 for GGG. With $v_{\rm GGG} =3.53~\rm km/s$ for the transverse sound wave, the phonon signal is expected and detected at about 26 GHz}. The simultaneously obtained microwave reflection spectrum is shown in Fig.~\ref{FIG_setup}(d). The two spectra taken at 119.7 mT and 0 mT are compared, where the latter represents the background antenna reflection. The difference (red shaded area in Fig.~\ref{FIG_setup}(d)) shows the absorbed microwave power by the magnetic system. A broad resonance spectrum composed of multiple sharp peaks is obtained. The spectral broadening is attributed to the spatially inhomogeneous magnetic precession over the large detection area covered by the antenna.


With the capability of locally probing the dynamic magnetization, we now demonstrate the excellent sensitivity of BLS to the formation of magnon-polarons for out-of-plane magnetized configuration by applying $H_z$. Here the signal-to-noise ratio in the BLS measurement is low because BLS is less sensitive to the in-plane dynamic magnetization \cite{hamrle2010analytical}. To compensate for this, the BLS spectrum is taken at the middle of the antenna to maximize the torque by $h_{\rm rf}$. Figure~\ref{FIG_compare}(a) shows the integrated BLS intensity as a function of $H_z$ and excitation frequency. We clearly see strongly reduced intensities of down to 90\% at each phonon frequency indicated by the red dotted lines. These are regularly placed at $f_n=nv_{\rm GGG}/[2(d+s)]$, where $n$ is an integer, $v_{\rm GGG}$ is the transverse phonon velocity in GGG, $d$ and $s$ are the YIG and GGG thickness values, respectively. With $v_{\rm GGG}=3.53~\rm km/s$ \cite{ye1991magnetoelastic}, we obtain a frequency spacing of $\Delta f \equiv f_n-f_{n-1}=5.34~\rm MHz$, close to the measured frequency spacing of 5.24 MHz. This spacing is enlarged compared to the previous works that used thicker GGG substrates (see Appendix E). The phonon-induced gap is about $\Omega_{\rm pp}/(2\pi)=3.1~\rm MHz$, where $\Omega_{\rm pp}$ is slightly larger than the actual coupling strength $\Omega$ by $\Omega_{\rm pp}\approx \Omega [1 + (\eta_m\eta_a)/\Omega^2]$. With $\eta_a/(2\pi)=0.45~\rm MHz$ and $\eta_m/(2\pi)=1.5~\rm MHz$ extracted from the fit shown as the purple solid line in Fig.~\ref{FIG_compare}(b), we obtain $\Omega_{\rm out}/(2\pi)=2.9~\rm MHz$.

Based on the picture illustrated in Fig.~\ref{FIG_illlus}, we expect a reduced magnon-phonon coupling for the in-plane magnetized configuration, where lattice vibrations become linear. Figure~\ref{FIG_compare}(c) shows the spectra obtained upon applying the in-plane magnetic field $H_x$ at a similar frequency range to that of Fig.~\ref{FIG_compare}(a). The phonon-induced gap of $\Omega_{\rm pp}/(2\pi)$ is reduced to $1.9~\rm MHz$ (Fig.~\ref{FIG_compare}(c)). With $\eta_a/(2\pi)=0.4~\rm MHz$ and $\eta_m/(2\pi)=1.4~\rm MHz$, we obtain $\Omega_{\rm in}/(2\pi)=1.5~\rm MHz$. This shows that there is a factor of 2 enhancement in $\Omega$ for the out-of-plane magnetized case. In terms of $\mathscr{C}$, this represents a factor of 4 improvement. In addition, there is no sign of coupling between longitudinal phonons and magnons, which would give a larger frequency spacing \footnote{Longitudinal phonons have $v^{\rm long}_{\rm GGG}=6.4~\rm km/s$ \cite{kleszczewski1988phonon}, which would lead to 9.7 MHz frequency spacing, not visible in Fig.~\ref{FIG_compare}(c).}, an experimental evidence that the phonons are indeed transverse-linearly polarized for the in-plane magnetized case \cite{streib2018damping}. Additionally, it is worth noting that similar acoustic damping values of $\eta_a/(2\pi)=0.4~\rm MHz$  were obtained for both the out-of-plane and in-plane field cases. This suggests a close proximity in the damping rates between circularly polarized and linearly polarized phonons.


Next, we discuss experiments with different phonon wavelengths by changing the central field, $H_x^{\rm c}$ for an in-plane magnetized case. Figure~\ref{FIG_freqsweep}(a-e) shows the field and frequency-dependent BLS spectra at different central phonon mode indices indicated by the red arrows. The integrated BLS intensities are plotted in the first column of Fig.~\ref{FIG_freqsweep}, where the color scales are normalized to the maxima. The magnon resonances broaden, and the phonon dips reduce with increasing $H_x^{\rm c}$. The absolute BLS intensity significantly drops for higher $H_x^{\rm c}$ (see the second and third columns of Fig.~\ref{FIG_freqsweep}) due to increased magnon damping, which is proportional to $f_{\rm rf}$ \footnote{We also observed an increased microwave insertion loss of about 3 dBm over the studied frequency range, which may account for an additional reduction of signal at high frequencies}. The green linecut is drawn over where the magnon phonon coupling is not visible, therefore pure magnetic resonances are obtained. The purple line cut is drawn to characterize the magnon-polarons.

To perform quantitative analysis, we use the two coupled oscillator model described as follows \cite{an2020coherent}:

\begin{equation} 
\begin{split}
(\omega-\omega_m+i\eta_m)m^+ & =\Omega u^+/2 +\kappa h^+,\\
(\omega-\omega_a+i\eta_a)u^+ & =\Omega m^+/2,
\end{split}
\label{eq1}
\end{equation}

where $m^+/u^+$ are circularly polarized magnon/phonon amplitudes and $\eta_{m/a}$ are the magnetic/acoustic relaxation rates. $\omega_{m/a}$ are the magnetic/acoustic resonances, $\kappa$ is the coupling to the antenna, $h^+$ is the antenna field. By solving Eq.~\ref{eq1}, we obtain analytic expressions for $m^+$ and $u^+$ given by

\begin{equation}
\begin{split}
m^+ = -\frac{4 \, \kappa(\omega -\omega_a+i\eta_a) \, h^+}{\Omega^2-4(\omega-\omega_a+i\eta_a)(\omega-\omega_m+i\eta_m)},\\
u^+ = -\frac{2 \,\kappa\Omega \, h^+}{\Omega^2-4(\omega-\omega_a+i\eta_a)(\omega-\omega_m+i\eta_m)}.\\
\end{split}
\label{eq2}
\end{equation}


The measured BLS intensity is proportional to $\vert m^+\vert^2$ \cite{buchmeier2007intensity,birt2012deviation} that we use to fit the experimental data. The green lines in Fig.~\ref{FIG_freqsweep} correspond to the fit without phonon contributions, i.e., $\Omega=0$ and $\eta_a =0$ from which we extract $\omega_m$ and $\eta_m$. Then, we fit the case with magnon polarons (purple lines in Fig.~\ref{FIG_freqsweep}) to extract $\Omega$, $\omega_a$ and $\eta_a$. With these parameters, the complementary $|u^+|^2$ can be calculated based on Eq.~\ref{eq2}. We then estimate the relative power transferred to the phonon system by 

\begin{equation} 
P_a=\frac{\eta_a |u^+|^2}{\eta_m |m^+|^2+\eta_a |u^+|^2}.
\label{eq_Pa}
\end{equation}

At low $f_{\rm rf}$, $P_a$ becomes larger than 50\% (see the fourth column of Fig.~\ref{FIG_freqsweep}). The spectral integration, $\big<P_a\big>\equiv\int d\omega P_a$, becomes maximum near 9 GHz, where the magnon phonon coupling reaches the maximum (see Fig.~\ref{FIG_extract}(b)). It is important to note that the experimentally obtained spectral shape consistently aligns with $m^+$, confirming that our BLS method detects the magnon signal instead of phonons.

The frequency dependences of $\eta_m$, $\eta_a$, $\Omega$, and $\mathscr{C}$ are summarized in Fig.~\ref{FIG_extract}. $\eta_m$ follows the predicted linear frequency dependence well (green solid line in Fig.~\ref{FIG_extract}(a)) from which we extract $\alpha_{m}=1.1\times 10^{-4}$, which is close to literature value \cite{dubs2017sub}, and the inhomogeneous line broadening of $\Delta \eta_{m}/(2\pi)=0.58~\rm MHz$. A quadratic dependence on frequency is expected for acoustic damping (blue solid line in Fig.~\ref{FIG_extract}(a)) \cite{dutoit1974microwave}. We extract an acoustic relaxation rate of $\alpha_{a}=7.7\times 10^{-6}~\rm GHz^{-1}$ and an inhomogeneous contribution of $\Delta \eta_{a}/(2\pi)=0.16~\rm MHz$, which are in reasonable agreements with previously reported values \cite{kleszczewski1988phonon,schlitz2022magnetization}. The coupling strength for the in-plane magnetized case is given by (see Appendix B for derivation)

\begin{equation}
\frac{\Omega}{2\pi}  = B_{\rm eff}\sqrt{\frac{\gamma \omega_H}{2\pi^2\omega^2\rho M_s d(d+s)}} \Big( 1-\cos{\frac{\omega d}{v_{\rm YIG}}} \Big),
\label{coupling}
\end{equation}

where $\omega_H=\gamma \mu_0 H$, $\rho$ is the density of YIG, and $M_s$ is the saturation magnetization. $B_{\rm eff}$ represents the effective magnetoelastic coefficient. We use $\mu_0 M_s = 0.172~\rm T$, $\rho = 5100 \rm~kg/m^3$, $\gamma/(2\pi)=28~\rm GHz/T$, and the known transverse sound velocity $v_{\rm YIG}=3.84~\rm km/s$ \cite{clark1961elastic}. The expected coupling strength variation for in-plane (Eq.~\ref{coupling}) and out-of-plane (Eq.~\ref{couplingOut}) magnetized cases are shown with $B_{\rm eff} = 7\times 10^5~\rm J/m^3$ as solid lines in Fig.~\ref{FIG_extract}(b). We note that it works well for both in-plane and out-of-plane configurations. However this estimate of $B_{\rm eff}$ deviates from the theoretical value obtained with the known material parameters and assumed pinning free boundary conditions, $\overline{B}_{[1\bar{1}0]}=5.2 \times 10^5~\rm J/m^3$ (see Appendix B). We attribute the discrepancy to the several assumptions made in the calculations, i.e., unpinned spins at the boundaries, phonon properties assumed to be identical for the YIG film and GGG substrates, and neglected anisotropy fields \cite{polulyakh2021magnetoelastic}. This calls for detailed investigation on magnetoelastic properties in thin films. Additionally, it is important to note that while our study primarily focuses on the excitation of magnons with $k\sim 0$, there is a possibility that high-$k$ magnons could also interact with phonons. This raises intriguing questions about the variations in cooperativity for high-$k$ magnons. Finally the frequency variation of cooperativity is shown Fig.~\ref{FIG_extract}(c). The maximum $\mathscr{C}$ is achieved at lower frequency of about 5.5 GHz compared to 9 GHz for the optimal coupling strength due to the reduction of $\eta_a$ and $\eta_m$ with decreasing frequency. 


In conclusion, the local magnon-phonon coupling was investigated using an optical technique. Enhanced contrast was observed at the phonon resonances due to the reduced nonuniformity over the detection area. Furthermore, we demonstrated tunable magnon phonon coupling and determined optimal parameters for maximizng magnon phonon interconversion in a planar geometry. Stronger coupling strength with the out-of-plane magnetized configuration was observed, which agrees with the calculations. Our local sensing scheme and optimization of the interconversion may find application to the coherent quantum information processing.






\section*{Acknowledgments}
We thank Simon Streib for helpful discussions. This work was partially supported by the French Grants  ANR-21-CE24-0031 Harmony and the EU-project H2020-2020-FETOPEN k-NET-899646; the EU-project HORIZON-EIC-2021-PATHFINDEROPEN PALANTIRI-101046630. K.A. acknowledges support from the National Research Foundation of Korea (NRF) grant (NRF-2021R1C1C2012269) funded by the Korean government (MSIT). 

\begin{figure*}
\centering
\includegraphics[width=1\textwidth]{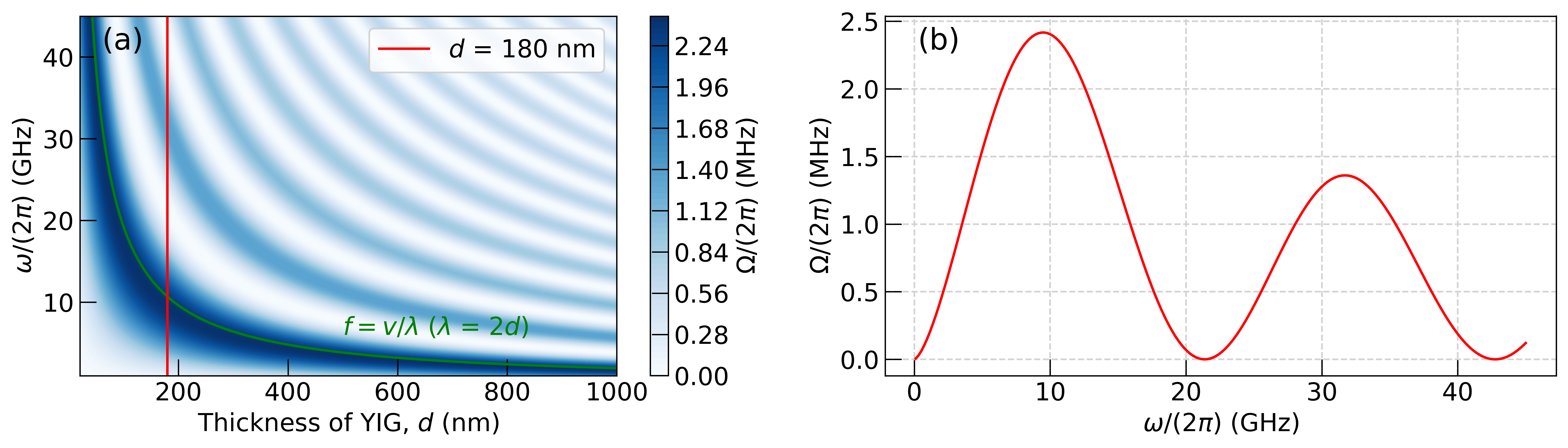}
\caption{(Color online) (a) Calculated coupling strength as a function of $d$ and $\omega$ based on Eq.~\ref{couplingOut} with the known material parameters. The green line indicates the maximum coupling at the half-wave condition. (b) Linecut along the red line in (a). }
\label{FIG_strength}
\end{figure*}

\section*{Appendix A : Magnetoelastic coupling strength 
in a thin magnetic film for the out-of-plane configuration}
\setcounter{equation}{0}
\renewcommand{\theequation}{A\arabic{equation}}

We derive an analytic form of coupling strength for out-of-plane magnetized case. We write the magnetic/acoustic equation of motion without damping and crystalline anisotropy terms \cite{comstock1963generation} :

\begin{equation}
\begin{split}
\omega m^+ = \gamma\mu_0(H-M_s) m^+ +\gamma \overline{B}_{[111]} \frac{\partial u^+}{\partial z}, \\
-\rho\omega^2 u^+ = C_{44}  \frac{\partial^2 u^+}{\partial z^2}  +\frac{\overline{B}_{[111]}}{M_s}\frac{\partial m^+}{\partial z}, \\
\end{split}
\label{eqA1}
\end{equation}
where $m^+$ and $u^+$ represent circularly polarized magnons and phonons. $\overline{B}_{[111]}$ here represents the effective magnetoelastic coefficient when the out-of-plane field is applied along the [111] direction and $C_{44}$ is elastic stiffness constant. We assume that $\textbf{M}$ and $\textbf{H}$ are parallel.

With $\omega_{m,\perp}=\gamma \mu_0 (H-M_s) $, $v^2=C_{44}/\rho$, and $\omega_a=vk$, Eq.~\ref{eqA1} can be written as

\begin{equation}
\begin{split}
(\omega -\omega_c) m^+ = \gamma \overline{B}_{[111]}\frac{\partial u^+}{\partial z}, \\
-(\omega -\omega_c) u^+= \frac{\overline{B}_{[111]}}{2\omega\rho M_s}\frac{\partial m^+}{\partial z}, \\
\end{split}
\label{eqA2}
\end{equation}

where we approximated the phonon part with $2\omega\approx\omega +\omega_a$ near the crossing point $\omega \approx \omega_c=\omega_{m,\perp}=\omega_a$. We consider the coordinate system shown in Fig.~\ref{FIG_setup}(a). The magnon profile is $m^+(z)=m^+_0\theta(-z)\theta(z+d)$ for unpinned spins at the boundaries, where $\theta(z)$ is the Heaviside step function. Phonon profile is $u^+(z)=u^+_0\cos \left [ n\pi(z+d) / (s+d) \right  ]$, where $n$ is the index of the phonon mode. This assumes that the phonon amplitude becomes maximum at the boundaries. Coupling strength represents the anticrossing gap, $\Omega = 2|\omega-\omega_c|$. By integrating over the whole volume, we obtain

\begin{equation}
\begin{split}
\frac{\Omega_{\rm [111]}^{\rm out}}{2}  \int_{-d}^s dz (m^+)^2 = \gamma \overline{B}_{[111]}\int_{-d}^s dz m^+\frac{\partial u^+}{\partial z} , \\
\frac{\Omega_{\rm [111]}^{\rm out}}{2} \int_{-d}^s dz (u^+)^2 = -\frac{\overline{B}_{[111]}}{2\omega\rho M_s}\int_{-d}^s dz u^+\frac{\partial m^+}{\partial z}, \\
\end{split}
\label{Eq8}
\end{equation}

The two equations are multiplied to yield

\begin{equation}
\begin{split}
\Big(\frac{\Omega_{\rm [111]}^{\rm out}}{2}\Big)^2 \frac{d(d+s)}{2}(m_0^+)^2(u_0^+)^2 \\
= -\frac{\gamma \overline{B}_{[111]}}{2\omega\rho M_s} \left(\int_{-d}^s dz m^+\dfrac{\partial u^+}{\partial z}\right)\left(\int_{-d}^s dz u^+\dfrac{\partial m^+}{\partial z}\right),
\end{split}
\end{equation}

where we applied the spatial integration. The integral on the right side can be evaluated using the properties of Heaviside step function and delta function, i.e., $\delta(z)=\partial\theta(z)/\partial z$ and $f(a)=\int dx f(x)\delta (x-a)$. Finally, we obtain

\begin{equation}
\Omega_{\rm [111]}^{\rm out}  = 2\overline{B}_{[111]}\sqrt{\frac{\gamma}{\omega\rho M_s d(d+s)}} \Big(1-\cos{\frac{n\pi d}{d+s}}\Big).
\label{couplingOut}
\end{equation}

From literature, we obtain a theoretical estimation of $\overline{B}_{[111]}= (2B_1+B_2)/3=4.64\times 10^5~\rm J/m^3$ \cite{comstock1963generation}. To visualize Eq.~\ref{couplingOut}, the coupling strength formula is plotted in Fig.~\ref{FIG_strength} with the parameters obtained from fitting experimental data.

\section*{Appendix B : Magnetoelastic coupling in a thin magnetic film for the in-plane field configuration}
\label{appendix:b}
\setcounter{equation}{0}
\renewcommand{\theequation}{B\arabic{equation}}

The equation of motion needs to be modified when field is applied along the in-plane $[1\bar{1}0]$ direction to take account of the change in $E_{\rm me}$ (see Appendix F). In this case, the modified equation of motion is given as \cite{polzikova2014magnetic}

\begin{equation}
\begin{split}
i\omega m_z=\gamma \mu_0 H m_y -\gamma \overline{B}_{[1\bar{1}0]}\frac{\partial u_x}{\partial z}\\
i\omega m_y=-\gamma\mu_0(H+M_s) m_z -\gamma \overline{B}_{[111]}\frac{\partial u_x}{\partial z}\\
-\rho \omega^2u_x=C_{44}\frac{\partial^2 u_x}{\partial z^2}+\frac{\overline{B}_{[111]}}{M_s}\frac{\partial m_z}{\partial z}-\frac{\overline{B}_{[1\bar{1}0]}}{M_s}\frac{\partial m_y}{\partial z},
\end{split}
\end{equation}

where $\overline{B}_{[1\bar{1}0]}$ is the magnetoelastic constants defined in Appendix F. The second equation for $m_y$ can be plugged in to yield 
\begin{equation}
\begin{split}
(\omega^2-\omega_{m,\parallel}^2) m_z= (\gamma\omega_H \overline{B}_{[111]}+i\omega\gamma \overline{B}_{[1\bar{1}0]})\frac{\partial u_x}{\partial z}\\
(\omega^2-\omega_a^2)u_x=-\Big(\frac{\overline{B}_{[111]}}{\rho M_s}+\frac{\overline{B}_{[1\bar{1}0]}(\omega_H+\omega_M)}{\rho M_s i\omega}\Big)\frac{\partial m_z}{\partial z},
\end{split}
\end{equation}

where $\omega_{m,\parallel} = \gamma \mu_0 \sqrt{H(H+M_s)}$ and $\omega_M=\gamma\mu_0 M_s$ Near the crossing point, we have $\omega\approx\omega_c\equiv\omega_{m,\parallel}=\omega_a$ and the above can be approximated as

\begin{equation}
\begin{split}
(\omega-\omega_c) m_z= \frac{(\gamma\omega_H \overline{B}_{[111]}+i\omega\gamma \overline{B}_{[1\bar{1}0]})}{2\omega}\frac{\partial u_x}{\partial z}\\
(\omega-\omega_c)u_x=-\Big(\frac{\overline{B}_{[111]}}{2\omega\rho M_s}+\frac{\overline{B}_{[1\bar{1}0]}(\omega_H+\omega_M)}{2\rho M_s i\omega^2}\Big)\frac{\partial m_z}{\partial z}
\end{split}
\end{equation}

We perform the integration outlined in appendix A. 

\begin{equation}
\begin{split}
\frac{\Omega_{[1\bar{1}0]}^{\rm in}}{2}\int_{-d}^s m_z^2dz= \frac{(\gamma \overline{B}_{[111]}\omega_H+i\gamma\omega \overline{B}_{[1\bar{1}0]})}{2\omega}\\
\times\int_{-d}^s\frac{\partial u_x}{\partial z}m_zdz\\
\\
\frac{\Omega_{[1\bar{1}0]}^{\rm in}}{2}\int_{-d}^s u_x^2 dz=\frac{-\overline{B}_{[111]}\omega+i\overline{B}_{[1\bar{1}0]}(\omega_H+\omega_M)}{2\rho M_s \omega^2}\\
\times\int_{-d}^s\frac{\partial m_z}{\partial z} u_x dz
\end{split}
\end{equation}

Multiplying the two yields

\begin{equation}
\begin{split}
\left(\frac{\Omega_{[1\bar{1}0]}^{\rm in}}{2}\right)^2\frac{d(d+s)}{2}= -\frac{\gamma}{4\omega^3\rho M_s}(\overline{B}_{[111]}\omega_H+i\omega \overline{B}_{[1\bar{1}0]})\\
 \times(-\overline{B}_{[111]}\omega+i\overline{B}_{[1\bar{1}0]}(\omega_H+\omega_M))\Big(1-\cos{\frac{n\pi d}{d+s}}\Big)^2
\end{split}
\end{equation}

The imaginary term concerns the size of gap, which is much smaller than the resonance frequency. Ignoring this, we obtain

\begin{equation}
\begin{split}
(\Omega_{[1\bar{1}0]}^{\rm in})^2= \frac{2\gamma}{\omega^2\rho M_sd(d+s)}(\overline{B}_{[111]}^2\omega_H\\
+\overline{B}_{[1\bar{1}0]}^2(\omega_H+\omega_M))
\times\Big(1-\cos{\frac{n\pi d}{d+s}}\Big)^2
\end{split}
\end{equation}

Finally the coupling strength is given by

\begin{equation}
\Omega_{[1\bar{1}0]}^{\rm in}= \overline{\overline{B}}_{[1\bar{1}0]} \sqrt{\frac{2\gamma \omega_H}{\omega^2\rho M_sd(d+s)}}\Big(1-\cos{\frac{n\pi d}{d+s}}\Big),
\label{couplingIn}
\end{equation}

where

\begin{equation}
\overline{\overline{B}}_{[1\bar{1}0]}=\sqrt{\overline{B}_{[111]}^2+\overline{B}_{[1\bar{1}0]}^2(1+\frac{\omega_M}{\omega_H})}.
\end{equation}

In our case, $\overline{B}_{[111]}=4.64\times 10^5~\rm J/m^3$, $\overline{B}_{[1\bar{1}0]}=-1.64\times 10^5~\rm J/m^3$, $\mu_0 M = 0.172~\rm T$, and $\mu_0H=0.157~\rm T$. Then $\overline{\overline{B}}_{[1\bar{1}0]}=5.21\times 10^5~\rm J/m^3$, which amounts to 12\% increase from $\overline{B}_{[111]}$.

The enhancement ratio is given by
\begin{equation}
\frac{\Omega_{[111]}^{\rm out}}{\Omega_{[1\bar{1}0]}^{\rm in}}=\frac{\overline{B}_{[111]}}{\overline{\overline{B}}_{[1\bar{1}0]}}\sqrt{\frac{2\omega}{\omega_H}}.
\label{couplingRatio}
\end{equation}

With $\omega = 2 \pi \times 6.4~\rm GHz$, we obtain the ratio of about 1.5. This value is smaller than the experimental result of 1.9. To further improve the agreement, one needs to take into account more accurate pinning conditions and anisotropy fields.

\begin{figure}
\centering
\includegraphics[width=0.48\textwidth]{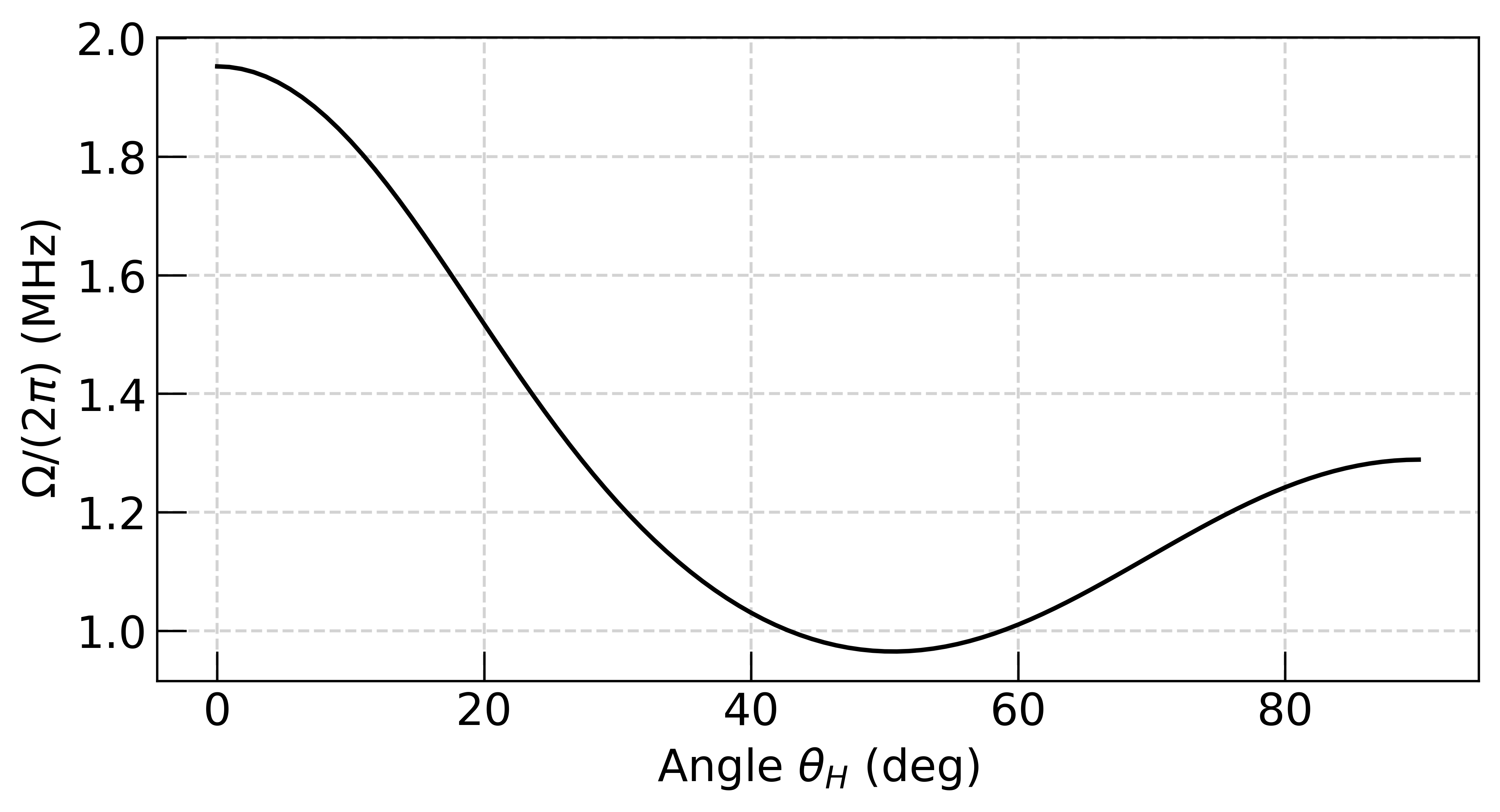}
\caption{(Color online) (a) Calculated angular dependence of coupling strength for the uniform magnetic precession and transverse phonons from Eq.~\ref{planarSchlomann}}
\label{FIG_Angle}
\end{figure}

\section*{Appendix C : Magnetoelastic coupling in a thin magnetic film for arbitrary magnetic field directions}
\label{appendix:c}
\setcounter{equation}{0}
\renewcommand{\theequation}{C\arabic{equation}}

\begin{figure*}[htp]
\centering
\includegraphics[width=1\textwidth]{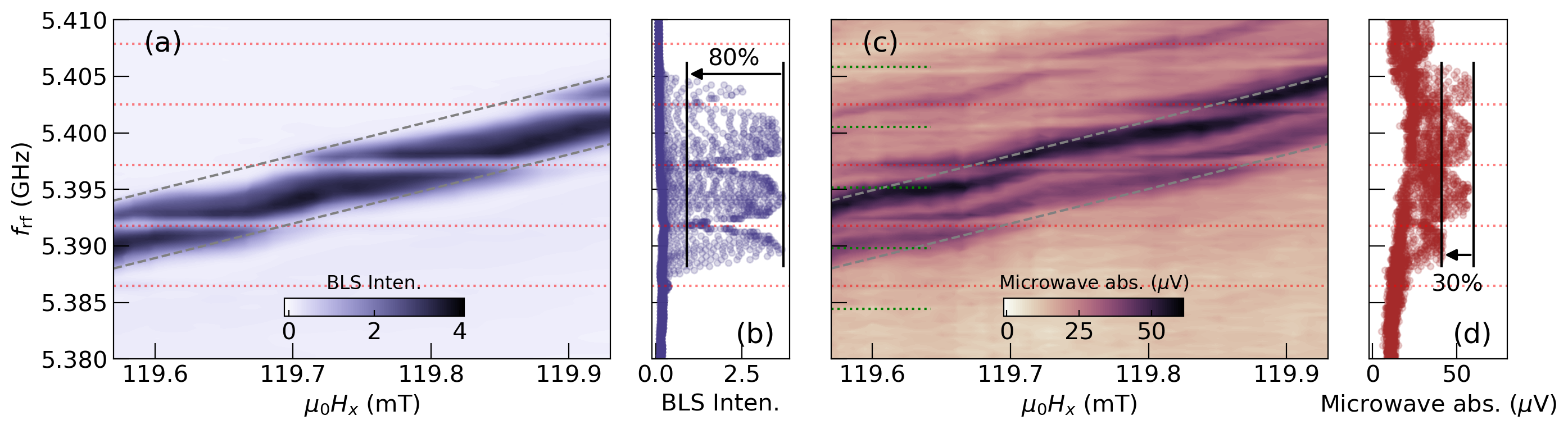}
\caption{(Color online) (a,b) BLS spectrum and (c,d) simultaneously obtained microwave absorption spectrum with field applied within the film plane. The red dotted lines represent phonon resonances. The green dotted lines show additional phonon lines only visible in the microwave absorption spectrum. The area between two diagnoal dashed lines compares the main magnon spectrum visible in BLS. Panels (b) and (d) show the frequency domain projection spectrum of the each density plot.}
\label{FIG_BLSvsdiode}
\end{figure*}

The coupling strength between propagating phonons and spin waves for arbitrary field directions has been derived by E. Schl$\rm\ddot{o}$mann \cite{schlomann1960generation}, which reads

\begin{equation}
\Omega = \sqrt\frac{\gamma F(\theta_H)}{\omega^2 \rho M_s}k,
\label{planarSchlomann}
\end{equation}

where $k$ represents the phonon wave vector and $\theta$ is the angle between the external magnetic field and the $z$ axis within the $xz$ plane. For simplicity, we assume that the directions of magnetization and external magnetic field are always parallel, which is indeed the case for sufficiently large external field. $F(\theta)$ is defined as

\begin{equation}
\begin{split}
F(\theta_H)=\overline{B}_{[111]}^2[\omega_H(\cos^2{2\theta_H}+\cos^2{\theta_H})\\
-\omega_M\cos^2{\theta_H}\cos{2\theta_H}(1+\cos{2\theta_H})]\\
+\overline{B}_{[1\bar{1}0]}^2\sin^2{\theta_H}[\omega_H\sin^2{\theta_H}
+\omega_M(\cos^4{\theta_H}-\cos{2\theta_H})].
\label{ftheta}
\end{split}
\end{equation}

For the field applied along the $z$ and $x$ direction, we obtain $F(0^{\circ})=2\overline{B}_{[111]}^2(\omega_H-\omega_M)$ and $F(90^{\circ})=\omega_H[\overline{B}_{[111]}^2+\overline{B}_{[1\bar{1}0]}^2(1+\omega_M/\omega_H)]$. This leads to the coupling strengths equivalent to Eq.~\ref{couplingOut} and Eq.~\ref{couplingIn} for the field applied along the $z$ and $x$ axis, respectively, when replacing $k$ by $\sqrt{\frac{2}{d(d+s)}}(1-\cos{\frac{n\pi d}{d+s}})$. The calculated angular dependence based on Eq.~\ref{planarSchlomann} is plotted in Fig.~\ref{FIG_Angle}.

\section*{Appendix D : Comparison between the BLS and FMR spectra}
\label{appendix:d}
\setcounter{equation}{0}
\renewcommand{\theequation}{D\arabic{equation}}

Here we provide a density plot of Fig.~\ref{FIG_setup}(d) as a function of field.  In BLS, phonons are visible with a better contrast due to its local detection. The intensity reduces by 80\% in Fig.~\ref{FIG_BLSvsdiode}(b) compared to the 30\% reduction in Fig.~\ref{FIG_BLSvsdiode}(d) at the phonon resonances. The sharper contrast in BLS is attributed to the reduced detection area leading to a more homogeneous condition. Also multiple magnon lines are visible in the FMR spectrum as shown in Fig.~\ref{FIG_BLSvsdiode}(c). In addition to the main phonon lines indicated by red dotted lines, there is also a set of phonon lines visible indicated by green dotted lines. 

\begin{figure}[b]
\centering
\includegraphics[width=0.5\textwidth]{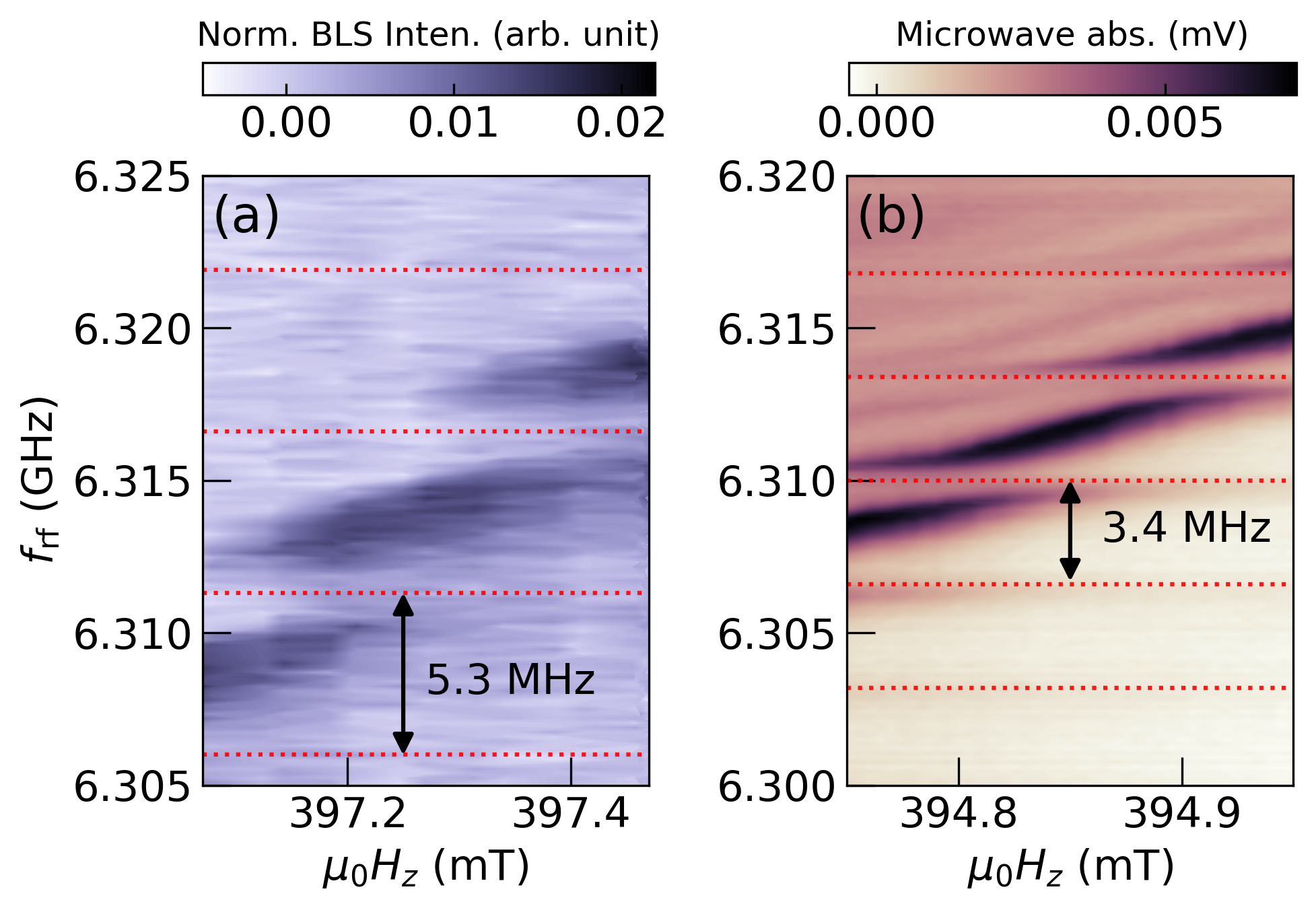}
\caption{(Color online) (a) BLS spectrum from 180 nm thick YIG on a 330 $\mu$m thick GGG. (b) Microwave absorption spectrum from 200 nm thick YIG on a 500 $\mu$m thick GGG. The phonon lines are positioned with smaller frequency spacing due to the thicker GGG substrate. Both spectra are taken with the field applied normal to the film surface. Red dotted lines represent the phonon resonances.}
\label{FIG_thicknessGGG}
\end{figure}

\section*{Appendix E : Frequency spacing for different thicknesses of GGG substrates}
\label{appendix:e}
\setcounter{equation}{0}
\renewcommand{\theequation}{E\arabic{equation}}

The thickness of GGG determines the phonon frequencies following $f_n=nv_{\rm GGG}/[2(d+s)]$, where $d$ and $s$ are the YIG and GGG thickness. With the $v_{\rm GGG}=3.53~\rm km/s$, we obtain $\Delta f=5.3~\rm MHz$ and $\Delta f=3.5~\rm MHz$ for 330 $\mu$m and 500 $\mu$m thick GGG substrates, respectively. These estimations are close to the measured frequency spacing shown in Fig.~\ref{FIG_thicknessGGG}.

\section*{Appendix F : Magnetoelastic coefficients for the [1$\bar{1}$0] direction}
\begin{figure}[t]
\centering
\includegraphics[width=0.3\textwidth]{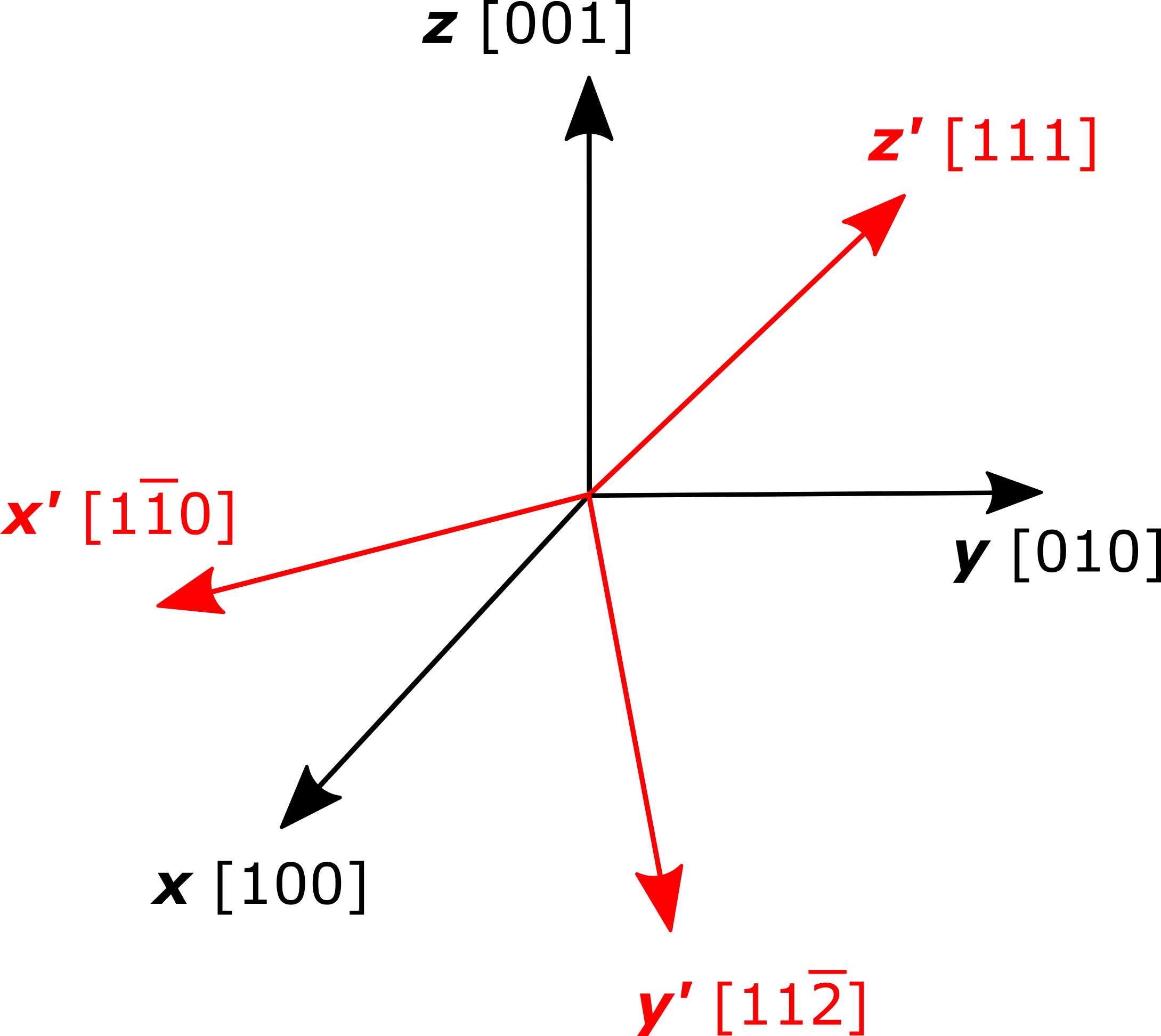}
\caption{(Color online) Coordinate transformation corresponding to Eq.~\ref{coordmat}.}
\label{FIG_coordtrans}
\end{figure}

\setcounter{equation}{0}
\renewcommand{\theequation}{F\arabic{equation}}
The value of $\overline{B}$ changes depending on the direction of magnetization with the respect to the crystallographic axis. If the system is fully isotropic then we have $B_1=B_2$. The difference between $B_1$ and $B_2$ would determines the variation of magnetoelastic energy with magnetization direction. The magnetoelastic energy then can be written as 

\begin{equation}
\begin{split}
E_{\rm me}=B_2(\sum_i m_i^2e_{ii}+\sum_{(ij)}m_i m_j e_{ij})\\
+(B_1-B_2)\sum_i m_i^2e_{ii},
\end{split}
\label{me_energy1}
\end{equation}

where $e_{ii}$ and $e_{ij}$ represents the strain tensor and the sum $(ij)$ goes over $(ij)=xy, yz, xz$. This form easily allows to extract $\overline{B}$ when magnetized along the [001] direction. The coefficients of $e_{ii}$ and $e_{ij}$ terms are $B_1$ and $B_2$, which represent the magnetoelastic constants for the longitudinal and transverse waves, respectively. When magnetized along the [111] direction, the transverse magnetoelastic coefficient is $\overline{B}_{[111]}=(2B_1+B_2)/3$ \cite{comstock1963generation}. In our experiments, the in-plane field is applied along the [1$\bar{1}$0] direction. To calculate the magnetoelastic energy in this case, one needs to perform a coordinate transform, where $x:[100]\rightarrow x':[1\bar{1}0]$, $y:[010]\rightarrow y':[11\bar{2}]$, and $z:[001]\rightarrow z':[111]$ (see Fig.~\ref{FIG_coordtrans}). The transformation matrix is given by

\begin{equation}
\mathscr{T}=
\begin{pmatrix}
1/\sqrt{2} & 1/\sqrt{6} & 1/\sqrt{3}\\
-1/\sqrt{2} & 1/\sqrt{6} & 1/\sqrt{3}\\
0 & -\sqrt{2}/\sqrt{3} & 1/\sqrt{3}
\end{pmatrix}.
\label{coordmat}
\end{equation}

Since only the second part of Eq.~\ref{me_energy1} concerns the anisotropic nature of magnetoelastic coupling, the first part should be invariant under the coordinate transform. The transformation of the second part can be written as 

\begin{equation}
\begin{split}
\sum_{i} m_i^2e_{ii}=\sum_{i,\nu',\mu',(\nu\mu)}  \mathscr{T}_{i\nu'}\mathscr{T}_{i\mu'}\mathscr{T}_{i\nu}\mathscr{T}_{i\mu}m_{\nu'}m_{\mu'}e_{\nu\mu}.
\end{split}
\label{me_energy_second}
\end{equation}

The magnetoelastic energy after this transformation has many terms. However most of them can be neglected in the first order approximation of dynamic magnetization. When the field is applied along the $x'$ direction, $m_{x'} m_{y'}$ and $m_{x'} m_{z'}$ are  the sole first order terms. Also we assume that the uniform magnetic precession excites the lattice. Therefore only $e_{x'z'}$ and $e_{y'z'}$ terms survive. The magnetoelastic energy therefore can be written as \cite{mason1965physical}

\begin{equation}
E^{[1\bar{1}0]}_{\rm me}\approx\overline{B}_{[111]} m_{x'}m_{z'} e_{x'z'}
+\overline{B}_{[1\bar{1}0]}m_{x'} m_{y'} e_{x'z'},
\label{me_energy3}
\end{equation}

where $\overline{B}_{[1\bar{1}0]}\equiv\ \sqrt{2}(B_1-B_2)/3$, which is about three times smaller than $\overline{B}_{[111]}$. The modification due to the nonzero $\overline{B}_{[1\bar{1}0]}$ is considered for the calculation described in Appendix B.

\bibliography{bib}

\begin{thebibliography}{61}
\bibitem [{foo()}]{footnote_fig3}%
  \BibitemOpen
  \href@noop {} {\bibinfo  {journal} {Higher power in the vertical field BLS experiments is necessary to enhance the low signal to noise ratio. The gap size remains constant within this power range}\ }\BibitemShut {NoStop}%
\end{thebibliography}

\clearpage


\end{document}